# Brain-wide connectome inferences using functional connectivity MultiVariate Pattern Analyses (fc-MVPA)

Alfonso Nieto-Castanon[1] [2]

[1] Department of Speech, Language, and Hearing Sciences. Boston University. Boston, MA. USA
[2] McGovern Institute for Brain Research. Massachusetts Institute of Technology. Cambridge, MA. USA

**Abstract**

Current functional Magnetic Resonance Imaging technology is able to resolve billions of individual functional connections characterizing the human connectome. Classical statistical inferential procedures attempting to make valid inferences across this many measures from a reduced set of observations and from a limited number of subjects can be severely underpowered for any but the largest effect sizes. This manuscript discusses fc-MVPA (functional connectivity Multivariate Pattern Analysis), a novel application of multivariate pattern analysis techniques in the context of brain-wide connectome inferences. The theory behind fc-MVPA is presented, and several of its key concepts are illustrated through examples from a publicly available resting state dataset, including an example analysis evaluating gender differences across the entire functional connectome. Last, Monte Carlo simulations are used to demonstrated this method's validity and sensitivity. In addition to offering powerful whole-brain inferences, fc-MVPA also provides a meaningful characterization of the heterogeneity in functional connectivity across subjects.

**Corresponding author**: Nieto-Castanon, Alfonso (alfnie@bu.edu)

**Introduction**

Functional connectivity Magnetic Resonance Imaging (fcMRI) is used to characterize the state (e.g. during rest or during individual cognitive tasks) of the human connectome, the set of all functional connections within the brain. In its most basic form, the entire human connectome state can be represented in a way that is only limited by the spatial resolution of the MRI acquisition sequence as a matrix of voxel-to-voxel functional connectivity values. Human connectome research is often motivated by the attempt to characterize similarities and discrepancies in these functional connectivity matrices across subjects or across experimental conditions, performing inferences that extrapolate form the limited data available in a study sample to properties of the human connectome in a larger population. Yet, this form of unconstrained **brain-wide connectome inferences** can suffer from a curse of dimensionality. A mass-univariate approach analyzing each functional connection between every pair of voxels in the brain may consist of over 60 *billion* individual statistical tests (the total number of pairwise functional connections among approximately 250,000 isotropic 2mm voxels within gray matter areas). This poses considerably difficulties. First, analytically, as an appropriate correction for multiple comparisons across this abundance of tests results in exceedingly low sensitivity. For example, simple Bonferroni or False Discovery Rate corrections (Benjamini and Hochberg 1995, Chumbley et al. 2010) would require at least one individual connection below $p<10^{-12}$ significance level in order to resolve FWE-corrected significance at the analysis level, limiting the sensitivity and applicability of these analyses. Second, computationally, as each voxel-to-voxel correlation matrix would require approximately 400Gb of memory or storage space for each individual subject and experimental condition of interest, making storing and working with these matrices extraordinarily demanding. And third, practically, as the potential wealth of information of voxel-to-voxel connectivity results makes reporting and interpreting the results of these analyses a significant challenge in itself.

Existing approaches have addressed these issues by either narrowing the focus of the analyses to connectivity with one or a few a priori seed areas (e.g. connectivity with amygdala) and then performing seed-based connectivity analyses (SBC), or by limiting the analysis units from voxels to larger parcels or Regions of Interest (ROIs) and then performing ROI-to-ROI connectivity analyses (RRC). The first approach (SBC) reduces the multiple-comparison problem by focusing on individual (or linear combinations of) rows of the functional connectivity matrices, disregarding all other possible effects beyond those involving at least one of these a priori seed areas. The main advantage of this approach is its simplicity, as it can take advantage of the same cluster-level inferential procedures that have been proved effective in standard analyses of functional activation, such as Gaussian Random Field theory inferences (Worsley et al. 1996), permutation/randomization analyses (Bullmore et al. 1999) or Threshold Free Cluster Enhancement (TFCE, Smith and Nichols, 2009). The main disadvantage of this approach is a high chance of potential false negatives, as other effects not involving the chosen seed areas may be missed.

The second approach (RRC) is able to perform brain-wide connectome inferences with sufficient sensitivity by drastically reducing the multiple-comparison problem, typically focusing on no more than a few hundred ROIs, often defining an entire parcellation of the brain (e.g. Harvard-Oxford atlas). In addition, specialized false positive control approaches, such as Functional Network Connectivity (Jafri et al. 2008), Network Based Statistics (Zalesky et al. 2010), Spatial Pairwise Clustering (Zalesky et al. 2012), or Threshold Free Cluster Enhancement (Smith and Nichols

2009) can be used to further increase the sensitivity of these analyses. ROI-to-ROI analyses, nevertheless, suffer from reduced spatial specificity arising from the a priori selection of parcels of interest, and their conclusions can be particularly sensitive to the choice of ROIs. Advances in connectivity-based parcelations (e.g. Gorbach et al. 2011) or subject-specific functional ROIs (e.g. Nieto-Castanon and Fedorenko, 2012) can be useful to partially alleviate these concerns.

An alternative approach uses Principal Component Analyses (PCA) or Independent Component Analyses (ICA) to assess differences in functional networks, or sets of functionally correlated areas, across groups (Calhoun et al. 2001). Similar to seed-based approaches, PCA and ICA are able to drastically reduce the multiple-comparison problem by focusing on individual networks, each comprising a group of functional-related areas, and then evaluating measures of within- and between- network connectivity. One advantage of this approach, compared to SBC, is that these networks do not need to be defined a priori and can be instead estimated directly from the functional data. In the context of brain-wide connectome inferences, nevertheless, these methods suffer from similar shortcomings as SBC, namely the potential of false negatives, where finer functional connectivity differences which are not present at the level of entire networks may be missed.

This paper discusses an alternative approach to the analysis of the brain-wide connectome at the resolution of individual voxels that overcomes the difficulties of brain-wide connectome analyses by using functional connectivity Multivariate Pattern Analysis (fc-MVPA). Like other MVPA approaches, fc-MVPA follows a general searchlight procedure (e.g. Kriegeskorte et al. 2006), considering separately for each voxel the entire multivariate pattern of functional connections between this voxel and the rest of the brain. Classical MVPA analyses (e.g. Norman et al. 2006) attempt to predict, from these or similar searchlight patterns, properties of the subjects (e.g. patients vs. controls) or properties of the experimental paradigm (e.g. pre- vs. after- intervention). These are often referred to as **backward models** (Haufe et al. 2014). fc-MVPA uses instead a **forward model** of the data, attempting to predict the shape of these searchlight patterns from subject and experimental paradigm information. In addition to being advantageous in terms of the interpretability of model weights, this approach also enables us to frame brain-wide connectome inferences in the context of the General Linear Model (GLM), one of the most widely used inferential statistical methods in neuroimaging. Specifically, this approach allows us to make classical statistical inferences about individual voxels in the brain regarding the *shape* of their functional connectivity patterns (e.g. is the shape of the functional connectivity pattern between a voxel and the rest of the brain different in patient vs. control subjects?), and then repeat these analyses across all voxels within the brain extending these inferences to the entire connectome (e.g. is the shape of functional connectivity patterns different in patient vs. control subjects anywhere in the brain?).

Fc-MVPA has been available as part of the CONN functional connectivity toolbox (Whitfield-Gabrieli and Nieto-Castanon, 2012) since 2012 and it has been already used to address a wide variety of research questions (e.g. Beaty et al. 2015, Flodin et al. 2016, Thompson et al. 2016, Amad et al. 2017, Beaty et al. 2017, Möller et al. 2017, Yankouskaya et al. 2017, Arnold Anteraper et al. 2019, Argyropoulos et al. 2019, Guell et al.2019, Morris et al. 2019, Multani et al. 2019, Schneider et al. 2019, Tortora et al. 2019, Argyropoulos et al. 2020, Guell et al. 2020, Kelly et al. 2020, Westfall et al. 2020, Katsumi et al. 2021, Mateu-Estivill et al. 2021, Shaw et al. 2021, Cahart et al. 2022, Eckstein et al. 2022, Fitzgerald et al. 2022, Smith et al. 2022, Walsh et al. 2022). Despite this, there has not been to date a rigorous description of this method's approach, validity, and motivation. This manuscript aims to correct that record by presenting a detailed description of fc-MVPA from the general perspective of other multivariate statistical inferential procedures, illustrating some of its key concepts through examples, and demonstrating its validity and sensitivity through simulations.

In the first section, a general framework to perform univariate and multivariate statistical inferences in the context of functional connectivity data is presented. This is followed by a section describing the specifics of the fc-MVPA approach, where some of its central applications, including brain-wide connectome inferences as well as the characterization of inter-subject heterogeneity, are further illustrated with examples. Finally, the last section presents simulation results demonstrating the method's validity, and discussing some of the main factors affecting sensitivity. All analysis examples in this manuscript are based on a publicly available resting state dataset (Cambridge 1000-connectomes dataset, n=198; www.nitrc.org/projects/fcon_1000), and all software and methods used are publicly available as part of SPM12 (RRID:SCR_007037, www.fil.ion.ucl.ac.uk/spm/software/spm12/) and CONN (RRID:SCR_009550, www.nitrc.org/projects/conn).

**Definition of functional-connectivity univariate and multivariate analyses (fc-MUA, fc-MVA)**

For any subject *n* in a study, and any pair of voxels *x* and *y*, we consider the values $r_n(x,y)$ characterizing the functional connectivity between these two voxels for this subject. Without loss of generality, we are going to consider $r_n(x,y)$ to represent Pearson correlation coefficients between these two voxels BOLD timeseries, but all of the discussion in this paper would equally apply to any arbitrary connectivity measure. In a study we desire to make inferences regarding properties of these connectivity measures in the population where the study subjects are being drawn from. For example, we may ask, based on our current study data, whether there are any connectivity differences when comparing patients to control subjects, or whether functional connectivity strength correlates with age, or whether it is modulated by some experimental condition. To that end we could use a forward model of the data defining for each individual connection a separate General Linear Model (GLM) of the form[1]:

---

functional connectivity Mass-Univariate Analyses (fc-MUA)

$$\forall x, y \quad r_n(x,y) = \boldsymbol{g}_n \cdot \boldsymbol{b}(x,y) + \varepsilon_n(x,y) \cdot \sigma(x,y) \quad (1)$$

Null hypothesis: $\quad \boldsymbol{C} \cdot \boldsymbol{b}(x,y) = 0$

---

where $\boldsymbol{g}_n$ is a predictor vector for each subject *n* characterizing known factors in our experimental design, such as group membership or behavioral measures (also known as the model *design matrix*), *b(x,y)* is an unknown vector of *regression coefficients*, estimated from the data and characterizing the effect of each modeled predictor on the outcome functional connectivity measures (e.g. the average connectivity strength within each group), *εₙ(x,y)* represents an error term sampled independently for each subject from a random Gaussian field with zero mean and unit variance (GLM asymptotic normality assumption), and $\sigma(x,y)$ is a variance term which depends on position *x* but is otherwise constant across subjects (GLM homoscedasticity assumption). The General Linear Model uses Ordinary Least Squares (OLS) to estimate the vector **b** from the data. After estimating these regressor coefficients, we can specify a null-hypotheses of the form $\boldsymbol{C} \cdot \boldsymbol{b}(x,y) = 0$ for any given between-subjects contrast **C** (e.g. a null hypothesis might evaluate whether functional connectivity differs between patients and controls) and use a classical hypothesis testing framework to evaluate this hypothesis. Hypothesis testing extrapolates from the observed properties of **b**, estimated only from our sample data, to the effects of the associated predictor terms in the larger population, allowing us to make valid inferences about any hypothesis in this larger population. Null-hypothesis are generally evaluated using a Likelihood Ratio Test based on Wilks Lambda distribution and associated *T*- and *F*-statistics (Nieto-Castanon, 2020). These hypotheses are tested separately for each pair of voxels *x* and *y*, resulting in a statistic parametric map *F(x,y)* of *T*- or *F*- statistics and associated p- values, characterizing the likelihood of our observations under the null hypothesis for every individual seed and target voxels.

---

[1] notation: in this document's equations we use regular fonts to refer to scalars, bolded lower case fonts for vectors, and bolded capital fonts for matrices

We refer to this approach as *mass-univariate* (fc-MUA) because it is based on a separate univariate test for each connection (for each pair of voxels *x* and *y*) in the entire brain-wide connectome. As mentioned before, one of the main difficulties with fc-MUA in the context of brain-wide connectome inferences is the extremely large number of connections evaluated (one for every pair of voxels) leading to the need for very strong multiple comparison corrections and reduced sensitivity to detect anything but the largest effects.

Compared to this mass-univariate approach, functional connectivity MultiVariate Analyses (fc-MVA) use a searchlight approach where each individual analysis focuses on one individual voxel-of-interest *x*, and perform an omnibus test of the connectivity between this voxel and the rest of the brain using a multivariate GLM analysis of the form:

---

functional connectivity MultiVariate Analyses (fc-MVA)

$$\forall x \quad \boldsymbol{r}_n(x) = \boldsymbol{g}_n \cdot \boldsymbol{B}(x) + \boldsymbol{\varepsilon}_n(x) \cdot \boldsymbol{\Sigma}(x) \qquad (2)$$

Null hypothesis: $\quad \boldsymbol{C} \cdot \boldsymbol{B}(x) \cdot \boldsymbol{P}(x) = \boldsymbol{0}$

---

The term $\boldsymbol{r}_n(x)$ in Equation 2 contains an entire map of connectivity values defined in vector form (each element of this vector contains the connectivity value $r_n(x, y)$ for a different target voxel *y*), fully characterizing the functional connectivity pattern for subject *n* between the seed-voxel *x* and the rest of the brain. *B(x)* is now an unknown predictors-by- voxels matrix of regression coefficients, $\boldsymbol{\varepsilon}_n(x)$ is a residual error vector sampled from a random multivariate Gaussian distribution with zero mean and unit variance, and $\boldsymbol{\Sigma}(x)$ is a voxels-by-voxels semipositive definite matrix characterizing the spatial covariance in functional connectivity patterns, which again may depend on position *x* but is otherwise constant across subjects. As before, the General Linear Model uses Ordinary Least Squares (OLS) to estimate the matrix of regressor coefficients *B* from the data. In the context of hypothesis testing, *C* and *P(x)* represent the between-subjects and between-voxels contrasts matrices, respectively, characterizing which aspects of the regression coefficients matrix we would like to evaluate. Any arbitrary hypothesis of the form $\boldsymbol{C}^t \cdot \boldsymbol{B}(x) \cdot \boldsymbol{P}(x) = \boldsymbol{0}$, may be evaluated separately for each searchlight voxel *x* using a statistical parametric map *F*(x), computed using a Satterthwaite approximation (Satterthwaite, 1946):

$$F(x) = \frac{b \; tr(\boldsymbol{H}(x))}{c \; tr(\boldsymbol{W}(x))} \sim F_{kc,kb} \qquad (3)$$

$$b \equiv N - rank(\boldsymbol{G})$$

$$c \equiv rank(\boldsymbol{G}\boldsymbol{C}^t)$$

$$k \equiv \frac{tr(\boldsymbol{W}(x))^2}{tr(\boldsymbol{W}^2(x))}$$

where *b* and *c* are error and hypothesis degrees of freedom, respectively, and the matrices *W* and *H* are the error and hypothesis sum of squares and products, respectively:

$$W(x) = P^t(x)\,(R(x) - GB(x))^t(R(x) - GB(x))\,P(x) \qquad (4)$$

$$H(x) = P^t(x)\,B^t(x)C^t(C(G^tG)^{-1}C^t)^{-1}CB(x)\,P(x)$$

$$B(x) = (G^tG)^{-1}G^tR(x)$$

$$R(x) \equiv [r_1^t(x)\ r_2^t(x)\ ...\ r_N^t(x)]^t$$

$$G \equiv [g_1^t\ g_2^t\ ...\ g_N^t]^t$$

Equation 3 results in a statistical parametric map *F(x)* with values that follow, under the null hypothesis, a standard *F* distribution with *kc* and *kb* degrees of freedom. This allows us to compute associated p- values characterizing the likelihood of our observations under the null hypothesis for every individual searchlight voxel.

The between-voxels contrast matrix $P(x)$ in Equation 2 serves to focus the analyses on a particular subspace of interest characterizing specific features of the functional connectivity maps $r_n(x)$. The choice of $P(x)$ affords great flexibility in the specific form of fc-MVA analyses that can possibly be implemented. For example, in the simplest possible scenario, we may choose $P(x)$ to be a constant one-dimensional projector, such as a unit-norm vector with positive weights over a single voxel or a small area, which would allow us to focus only in the connectivity with one a priori voxel or region of interest. Interestingly, in this scenario fc-MVA reduces exactly to a standard seed-based connectivity (SBC) analysis, producing statistical parametric maps *F(x)* that characterize the connectivity between the chosen voxel or area and the rest of the brain. In contrast, in perhaps the most general scenario, we may instead choose $P(x)$ to be also constant but now equal to the identity matrix, considering jointly and equally all target voxels. This allows us to simultaneously estimate and evaluate *any/all* aspects of the functional connectivity maps $r_n(x)$.

In between these two extrema, there are many reasonable alternatives. For example, a spatial basis $P(x)$ that would focus on low spatial-frequency components of connectivity profiles (e.g. Nieto-Castanon et al. 2003), one that would focus only on local connectivity with neighboring areas (e.g. a multivariate Local Connectivity measure), or one that would focus only on connectivity with all voxels within a fixed area (masked fc-MVA, e.g. connectivity with the cerebellum or any other large/heterogeneous area). In the next section we will discuss one particular form of fc-MVA analyses that is based on a data-driven choice of spatial basis $P(x)$ focusing on rich low-dimensional representations of arbitrary functional connectivity patterns.

**Definition of functional-connectivity multivariate pattern analyses (fc-MVPA)**

Functional connectivity multivariate *pattern* analyses (fc-MVPA) can be considered a particular case of functional connectivity multivariate analyses (fc-MVA), where the choice of spatial basis $P(x)$ attempts to achieve a balance between retaining high sensitivity to unknown or arbitrary effects while maintaining a good level of specificity to those features more representative of the data at hand. In particular, representative features in fc-MVPA are chosen to have maximal inter-subject variability and minimal overlap (i.e. orthogonal features). This is achieved by defining $P(x)$ explicitly as the right- orthogonal basis from a Singular Value Decomposition (SVD) of the connectivity matrix resulting from concatenating all of the maps $r_n(x)$ for a given *seed*-voxel *x* across all subjects:

$$R(x) \equiv [r_1^t(x)\ r_2^t(x)\ ...\ r_N^t(x)]^t = S(x) \cdot D(x) \cdot P^t(x)$$

$$S(x) \equiv [s_1^t(x)\ s_2^t(x)\ ...\ s_N^t(x)]^t = R(x) \cdot P(x) \cdot D^{-1}(x) \qquad (5)$$

where **S** and **P** are orthogonal matrices of left- and right- singular vectors, respectively, and **D** is a diagonal matrix of singular values sorted in decreasing order. The total number of singular vectors and values in Equation 5 is equal to the number of subjects *N*, but typically this dimensionality can be further reduced to only include the first few singular values and vectors that achieve a predefined predictive or descriptive target (e.g. those dimensions capturing on average 50% or more of the total covariance in the patterns of functional connectivity across subjects).

Conceptually, this particular choice of basis in fc-MVPA has a very important benefit, as the resulting *eigenpatterns*, defined as the columns of the resulting matrix $P(x)$, have a meaningful interpretation as those patterns that best characterize the observed heterogeneity across subjects in functional connectivity with an individual seed voxel. In particular the squared eigenvalues

$$\xi(x) \equiv diag\big(D^2(x)\big)/trace\big(R(x) \cdot R^t(x)\big) \tag{6}$$

represent the portion of the total inter-subject covariance $R(x) \cdot R^t(x)$ in connectivity maps that lies within the dimensions characterized by each individual eigenpattern, and by the nature of SVD these values are maximal (i.e. there is no other *k*-dimensional subspace containing a larger percentage of the total covariance of the data than the subspace spanned by the first *k* eigenpatterns, for any value *k*). In this context, the values $s_n(x)$, which we will refer in this manuscript as *eigenpattern scores*, define an optimal linear low-dimensional representation of the original data $r_n(x)$ for each subject, meaning that we can always linearly reconstruct the high-dimensional data $r_n(x)$ from its low-dimensional representation $s_n(x)$ with minimal error.

Mathematically, this approach is similar to functional PCA (Worsley et al. 1997) or to the group-level dimensionality reduction step in ICA (Calhoun et al. 2001) which helps reduce noise, simplify the analyses, and increase the interpretability of the resulting ICA components, but the main difference is that in fc-MVPA dimensionality reduction is performed separately for each individual seed-voxel *x*. Because the dimensionality reduction step in fc-MVPA is only tasked with characterizing the heterogeneity in functional connectivity patterns between one individual voxel *x* and the rest of the brain, while in PCA/ICA the dimensionality reduction step is tasked with simultaneously characterizing the heterogeneity in functional connectivity patterns between *all* pairs of voxels, the former can achieve a considerably more compact representation, where fewer components will explain a larger portion of that heterogeneity (as will be illustrated in the next section). In addition, the ability to obtain such low-dimensional characterization in a way that is specific to each anatomical location offers a considerably richer representation of inter-subject heterogeneity compared to other similar but global approaches, such as PCA or ICA.

In the context of brain-wide connectome analyses, using a simple change of basis allows us to simplify the fc-MVA multivariate general linear model and null hypothesis in Equation 2. In particular, by right-multiplying Equation 2 by the matrix $P(x) \cdot D^{-1}(x)$ we arise to an equivalent lower-dimensional fc-MVPA general linear model and hypothesis of the form:

functional connectivity MultiVariate Pattern Analyses (fc-MVPA)

$$\forall x \quad s_n(x) = g_n \cdot \widetilde{B}(x) + \varepsilon_n(x) \cdot \widetilde{\Sigma}(x) \tag{7}$$

Null hypothesis: $\quad C \cdot \widetilde{B}(x) = 0$

This is exactly the same model as in Equation 2 but expressed only within a lower-dimensional subspace represented by the eigenpattern scores $s_n(x)$, instead of the original higher-dimensional connectivity maps $r_n(x)$. In this context, the eigenpattern scores $s_n(x)$ represent what has also been referred to as Multivariate Connectivity maps (MCOR, Nieto-Castanon 2020), a voxel-specific low-dimensional multivariate representation of the pattern of functional connectivity between a voxel and the rest of the brain. Similarly, $\widetilde{B}(x)$ and $\widetilde{\Sigma}(x)$ in Equation 7 are also equal to their Equation 2 counterparts simply projected onto the subspace defined by $P(x)$. The reduced dimensionality allows us to simplify the computational implementation of these analyses considerably. For example, the eigenpattern scores $s_n(x)$ can now be simply stored as multiple whole-brain volumes, with one volume or image per component and per subject, and shared across multiple second-level analyses. This is in contrast with the considerably larger vectors $r_n(x)$ which cannot be easily stored (e.g. one whole-brain volume per subject *and per target voxel y*). In addition, the eigenpattern scores $s_n(x)$ are defined independently of the predictor vectors $g_n$, so not only they offer a model-free characterization of the heterogeneity in the data, but the same eigenpattern scores can also be used in multiple different group-level analyses of the same data. Last, the reduced dimensionality of Equation 7 also allows the covariance $\widetilde{\Sigma}^2(x)$ across eigenpattern scores to be fully estimable from limited number of samples, whereas the original covariance $\Sigma(x)$ across voxels in Equation 2 could very rarely be so estimated with full rank. Because of this, the effect of the within-subjects covariance in the resulting null hypothesis *F* statistics at each individual searchlight voxel does not need to be approximated (e.g. using Satterthwaite approximation as in Equation 3), allowing us to rely instead on a more sensitive Likelihood Ratio statistic (LRT) of the form (Rao, 1951):

$$F(x) = \frac{d}{ac} \cdot \frac{1 - \lambda^{1/e}}{\lambda^{1/e}} \sim F_{ac,d} \tag{8}$$

$$\lambda = \frac{|W|}{|W + H|}$$

$$a \equiv rank(P(x))$$

$$b \equiv N - rank(G)$$

$$c \equiv rank(GC^t)$$

$$d \equiv \left(b - \frac{a - c + 1}{2}\right) e - \frac{ac}{2} + 1$$

$$e \equiv \sqrt{\frac{a^2 c^2 - 4}{a^2 + c^2 - 5}}$$

where $\lambda$ is Wilks' Lambda statistic, *a* is the number of selected eigenpatterns (1<a<b), *b* is the error degrees of freedom, and *c* is the hypothesis degrees of freedom. As before, the resulting *F(x)* values follow, under the null hypothesis, a standard *F* distribution with *ac* and *b* degrees of freedom, for each individual searchlight voxel *x*.

To summarize, Figure 1 illustrates the idealized fc-MVPA procedure. For every searchlight voxel *x* we first compute the functional connectivity maps $r_n(x)$ between this voxel and the rest of the brain for every individual subject (Figure 1 top left), and use Equation 5 to compute a reduced set of eigenpattern scores $s_n(x)$ best characterizing relevant spatial features of these maps across subjects (represented in Figure 1 top right). Once each subject functional connectivity profiles are represented in terms of their lower-dimensional associated eigenpattern scores $s_n(x)$, group-level functional connectivity analyses proceed normally by entering these scores into a standard General Linear Model (Equation 7). This model evaluates at this searchlight location *x* any hypothesis of the form

$C \cdot \widetilde{B}(x) = 0$ using LRT (Equation 8), allowing us to make inferences about the shape of the functional connectivity maps that these scores represent. Last, this procedure is then simply repeated for each searchlight voxel *x*, sequentially constructing a statistical parametric map *F(x)* characterizing the results of this inferential procedure across the entire brain.

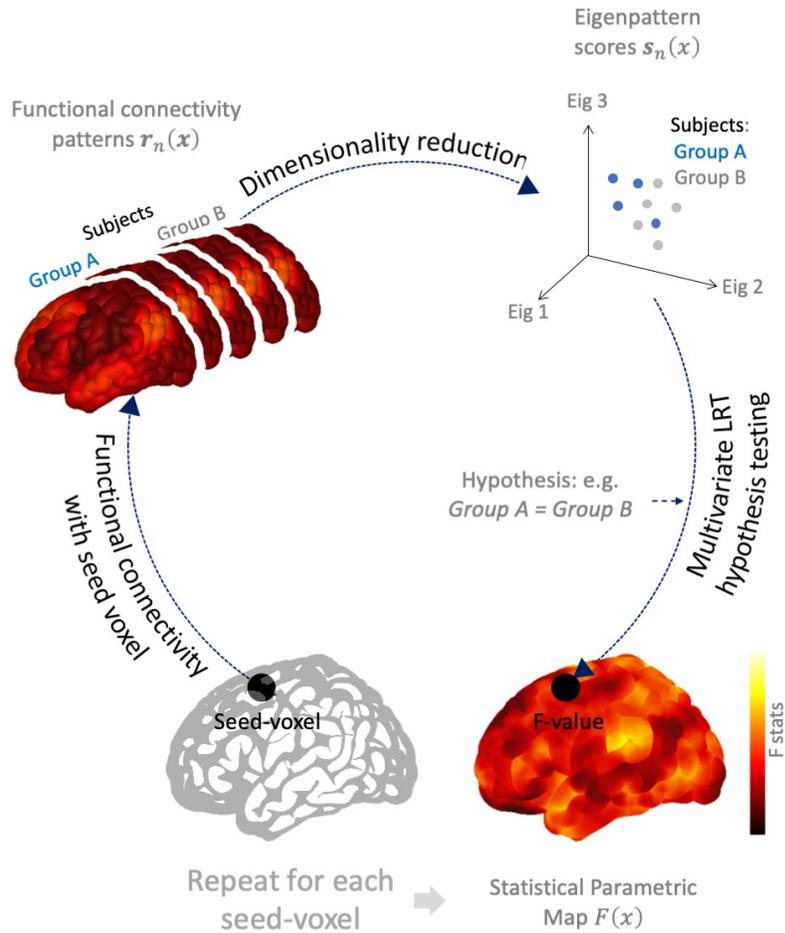

**Figure 1. Schematic representation of functional connectivity multivariate pattern analyses (fc-MVPA).** For each voxel, fc-MVPA analyses compute the functional connectivity maps between this seed/source voxel and the entire brain (Top-left; $r_n(x)$ in Equation 2) separately for each individual subject. Each subject functional connectivity map is then characterized by a lower dimensional eigenpattern scores (dots in top-right graph; $s_n(x)$ in Equation 5). This representation is chosen in a way that captures as well as possible the observed voxel-specific variability in functional connectivity maps across subjects. A multivariate test is then performed on the resulting lower-dimensional eigenpattern scores to ascertain potential between- or within- subjects effects of interest (e.g. differences between subjects or between conditions in functional connectivity at the original seed/source voxel). This process is then repeated for every source voxel to identify regions that show brain-wide between- or within- subjects differences in functional connectivity.

The general fc-MVPA procedure may be seen as computationally prohibitive, particularly for whole-brain analyses using relatively small voxel sizes (e.g. isotropic 2mm voxels), since the computational load scales quadratically with the total number of voxels under consideration, it appears to require the computation of entire voxel-to-voxel connectivity matrices, and it effectively performs close to 200,000 whole-brain PCA analyses (one per seed voxel) characterizing the inter-subject heterogeneity of seed-based connectivity maps. Despite this, there are several mathematical tricks that can be used to reduce by several orders of magnitude the complexity of the necessary computations. In particular, in Appendix I we describe how to more efficiently compute the eigenpattern scores $s_n(x)$ directly from the original BOLD timeseries in a way that will instead scale only linearly with the number of voxels, and without the need at any point to compute or store entire voxel-to-voxel connectivity matrices. In the analysis examples below we use this approach to efficiently compute fc-MVPA analyses on hundreds of subjects with minimal computation effort.

**Fc-MVPA brain-wide connectome inferences, interpretation and examples**

Group-level analyses of the eigenpattern scores $s_n(x)$ enable statistical inferences at the level of individual searchlight voxels evaluating the form or shape of the connectivity patterns with each voxel. In particular, for any individual hypothesis (e.g. group A = group B) the fc-MVPA procedure will produce a statistical parametric map $F(x)$ evaluating that hypothesis separately at each individual searchlight voxel *x*. In order to test brain-wide connectome hypotheses it is still necessary to control the resulting maps $F(x)$ for multiple comparisons across the total number of tests evaluated (one test per voxel). Fortunately, this can be done using the same non-parametric cluster-level inferential procedures that are common in standard analyses of functional activation, such as cluster-mass or TFCE statistics based on permutation / randomization analyses (Bullmore et al. 1999; Smith and Nichols, 2009, Winkler et al. 2014). These approaches allow us to compute statistics as well as associated family-wise error corrected p-values for individual **clusters** of contiguous searchlight voxels in the statistical parametric map $F(x)$, supporting cluster-level inferences with meaningful false positive control (e.g. controlling the likelihood of observing or more false positive clusters across the entire brain below 5%, for a family-wise control procedure, or controlling the rate of false positive clusters below 5% among all significant clusters, for a False Discovery Rate control procedure).

Another important choice that remains when using fc-MVPA in the context of brain-wide connectome inferences is to select the number of eigenpattern scores used to represent functional connectivity at each voxel. As it will be discussed in more detail in the *Simulations* section below, there is no "correct" choice of this parameter, and fc-MVPA inferences remain valid for all possible values of this parameter. Regarding sensitivity of fc-MVPA inferences, choosing a low number of eigenpatterns can be expected to improve sensitivity to detect relatively large or widespread effects of interest, while choosing a higher number of dimensions can improve our ability to detect relatively smaller or marginal effects. A reasonable balance is to scale the number of dimensions with the dataset size (e.g. suggested 5:1, 10:1, or 20:1 ratios between the number of subjects in the analysis and the number of eigenpattern scores retained, Vittinghoff et al. 2006, Concato et al. 1995, Harrell et al. 1996), in order to maintain a reasonable sensitivity to identify large effects in small samples, and comparatively finer details in the analysis of larger samples.

Last, regarding the interpretation of fc-MVPA results, when reporting statistical inferences from fc-MVPA brain-wide connectome analyses, those inferences should be if possible framed regarding the **patterns** of connectivity between each voxel or cluster and the rest of the brain. For example, when using fc-MVPA to evaluate the difference in connectivity between two groups of subjects, if the fc-MVPA procedure above produces one supra-threshold cluster with corrected significance level below p<.05 that should be interpreted as indicating that the pattern of connectivity between this cluster and the rest of the brain is (significantly) different between the two groups. The fc-MVPA method does not afford further spatial specificity in the resulting statistical inferences, but it is still useful to report measures of effect-size characterizing the patterns of connectivity within each individual significant cluster, as a way to suggest possible interpretations and future analyses. We recommend reporting the effect-sizes $\boldsymbol{h} \equiv \boldsymbol{c} \cdot \widetilde{\boldsymbol{B}}$ within each significant cluster $\Omega$ separately for each meaningful between-subjects contrast $\boldsymbol{c}$ (e.g. individual rows of the contrast matrix $C$). Effect sizes can be reported as a vector of eigenpattern weights ($\boldsymbol{h}_{\text{eig}}(\Omega)$), or as a whole-brain projected map ($\boldsymbol{h}_{\text{map}}(\Omega)$):

fc-MVPA effect-sizes at location $\Omega$

$$\boldsymbol{h}_{\text{eig}}(\Omega) \equiv \int_{x \in \Omega} \boldsymbol{c} \cdot \widetilde{\boldsymbol{B}}(x) \, dx \tag{9}$$

$$\boldsymbol{h}_{\text{scores}}(\Omega) \equiv \int_{x \in \Omega} \boldsymbol{c} \cdot \widetilde{\boldsymbol{B}}(x) \cdot \boldsymbol{S}^t(x) \, dx = \int_{x \in \Omega} \boldsymbol{h}_{eig}(x) \cdot \boldsymbol{S}^t(x) \, dx$$

$$\boldsymbol{h}_{\text{map}}(\Omega) \equiv \int_{x \in \Omega} \boldsymbol{c} \cdot \widetilde{\boldsymbol{B}}(x) \cdot \boldsymbol{D}(x) \cdot \boldsymbol{P}^t(x) \, dx = \int_{x \in \Omega} \boldsymbol{h}_{scores}(x) \cdot \boldsymbol{R}(x) \, dx$$

The effect-size measure $\boldsymbol{h}_{\text{eig}}(\Omega)$ is a vector, with one element per eigenpattern, estimated separately at each location $\Omega$. It represents the effect-size of a group-level analysis contrast of interest $\boldsymbol{c} \cdot \widetilde{\boldsymbol{B}}(x)$ evaluated separately for each individual eigenpattern (columns of $\widetilde{\boldsymbol{B}}$). For example, if the group-level analysis was a two-sample t-test comparing connectivity between two subject groups, then the n-th element in the $\boldsymbol{h}_{\text{eig}}(\Omega)$ effect-size vector will evaluate what is the difference in the nth-eigenpattern scores at location $\Omega$ between these two groups.

Similarly, and perhaps more directly interpretable, the effect-size measure $\boldsymbol{h}_{\text{map}}(\Omega)$ represents the same contrast but now evaluated separately at each individual voxel. In the example above, the value of $\boldsymbol{h}_{\text{map}}(\Omega)$ at a particular voxel will represent the difference in functional connectivity between $\Omega$ and this voxel between the two subject groups analyzed. It should be noted that a very similar whole-brain projected map of effect-sizes $\boldsymbol{h}_{map}(\Omega)$ can also be computed from the voxel-level effect-sizes of a post hoc analysis that would evaluate the same group-level model as in Equation 7 but this time focusing on the seed-based connectivity maps (SBC) associated with seeds defined from each individual significant cluster $\Omega$. As in any post-hoc analysis, p-values derived from these SBC post-hoc analyses will be partially inflated due to selection bias and should not to be used to make secondary inferences regarding individual connections within the reported patterns. Despite this limitation, post-hoc SBC analyses on the same dataset offer a simple and perfectly valid alternative approach for reporting fc-MVPA effect sizes within each significant cluster, while, when performed on an independent dataset, also offering a natural method to further probe specific aspects of these connectivity patterns. For those interested, the resulting $\boldsymbol{h}_{\text{map}}(\Omega)$ effect-sizes following this approach can be shown to be equal to those derived from Equation 9 in the limit when the number of eigenpatterns retained equals the total number of subjects in the study, making the variable $\boldsymbol{h}_{\text{scores}}(\Omega)$ simplify to a constant vector independent of the location $\Omega$:

$$\boldsymbol{h}_{\text{scores}}^{\infty}(\Omega) = \boldsymbol{c} \cdot (\boldsymbol{G}^t \cdot \boldsymbol{G})^{-1} \cdot \boldsymbol{G}^t \tag{10}$$

As an illustration of fc-MVPA brain-wide connectome inferences, we analyzed gender differences in resting state functional connectivity using the Cambridge 1000-connectomes dataset (n=198, see Appendix II for this dataset preprocessing and fc-MVPA analysis details). The question that these analyses ask is whether there are *any* differences across the entire voxel-to-voxel functional connectome between male and female subjects. To answer this question, we performed fc-MVPA analyses focusing on the first 10 eigenpatterns (an approximate 20:1 subjects to eigenpatterns ratio), entering the corresponding eigenpattern scores into a second-level group analysis evaluating a multivariate ANCOVA test with gender as a between-subjects factor and subject motion (average framewise displacement) as a control variable. The resulting statistical parametric maps were thresholded using Threshold Free

Cluster Enhancement (TFCE, Smith and Nichols, 2009, with default H=1, E=0.5 values) at a family-wise error corrected 5% false positive level.

The results, shown in Figure 2 show a large number of areas with significant gender-related differences in connectivity (p-FWE<.05, shown as yellow and black areas in center image). Given the abundance of areas showing significant gender effects, for illustration purposes we focused our description only on a subset of cortical regions showing some of the strongest effects (TFCE>200; p-FWE<.001, highlighted in black in Figure 2 center image). For each cluster in this reduced subset, we computed effect-size maps $h_{map}(\Omega)$ characterizing the pattern of gender-related differences in connectivity with each cluster, and these patterns are displayed in Figure 2 as a circular array of brain displays.

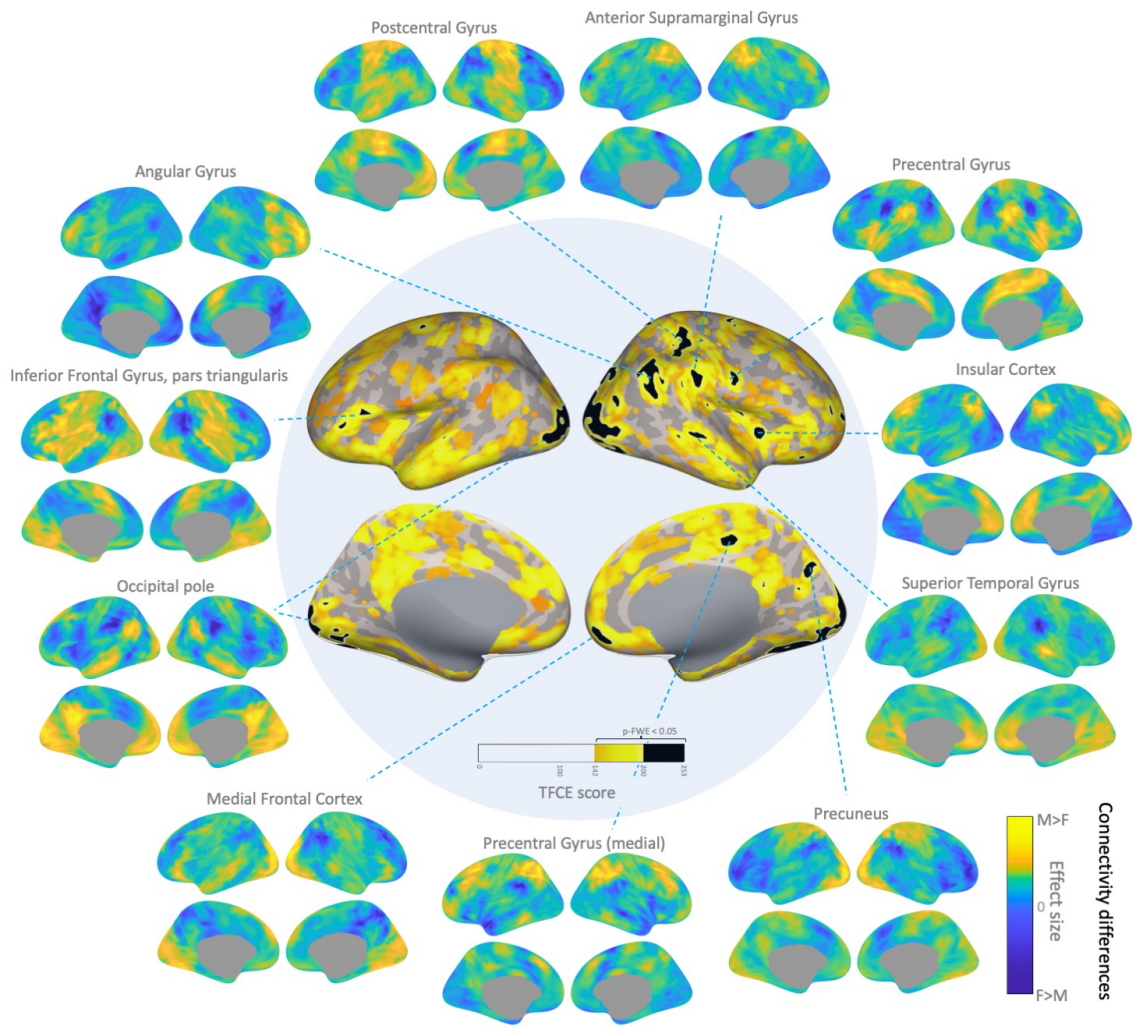

**Figure 2. Fc-MVPA results evaluating gender-related differences in connectivity**. Central figure shows left- and right- hemisphere medial (bottom) and lateral (top) views of the main fc-MVPA results showing areas with significant gender-related differences in functional connectivity (highlighted in yellow and black, TFCE statistics p-FWE<0.05). Among all significant results a reduced subset showing some of the strongest effects are highlighted in black, and the effect-sizes within these areas (pattern of differences in connectivity with each area between male and female subjects) are shown in the additional circular plots (yellow indicating higher connectivity in male compared to female subjects, and blue indicating higher connectivity in female compared to male subjects)

Some of the strongest effects were visible in bilateral Occipital Pole visual areas. A left hemisphere cluster centered at MNI coordinates (-22,-94,+4) mm showed a pattern of increased connectivity with Default Mode Network (DMN) and increased anticorrelations with Salience Network (SN) areas in male subjects (see Figure 2 Occipital Pole plot). A similar pattern (not shown) was present in another cluster at right hemisphere Occipital Pole areas (+28,-82,+2). Similarly, there were significant gender effects in several DMN areas, such as Medial Prefrontal Cortex (+6,+54,-12) and Precuneus (+18,-72,+32), showing a similar pattern of stronger connectivity with visual and sensorimotor areas (shown in yellow in Figure 2 Medial Frontal Cortex and Precuneus plots) in male subjects compared to stronger connectivity (weaker anticorrelations) with SN and attention areas in female subjects (shown in blue in same plots).

In the left hemisphere, Inferior Frontal Gyrus pars triangularis (-54,+22,+14) showed a pattern of stronger connectivity in female subjects with frontoparietal network areas and with Inferior Temporal Cortex (shown in blue in Figure 2 Inferior Frontal Gyrus plot).

In the right hemisphere, there were a cluster of regions in the Temporal Parietal Occipital Junction that also showed strong gender-related differences in connectivity. Lateral superior Postcentral Gyrus (+36,-32,+48) showed a mixed pattern of increased integration with other Dorsal Attention areas in female subjects, compared to increased connectivity with Central Sulcus, including somatosensory and motor areas in male subjects. A cluster in superior Angular Gyrus centered at coordinates (+50,-54,+42) mm showed increased integration with medial prefrontal and posterior cingulate areas in female subjects, and increased integration with lateral prefrontal and reduced anticorrelations with insular areas in male subjects. Relatedly, a cluster in mid Insular Cortex (+38,+10,+2) showed a similar pattern of higher connectivity with angular Gyrus and other DMN areas in male subjects. Another relatively proximal cluster in Anterior Supramarginal Gyrus (+60,-26,+30) also showed increased local associations with superior postcentral areas in male subjects. Posterior Superior Temporal Gyrus (+48,-26,-2) showed a pattern of higher connectivity (mixed with reduced anticorrelations) with frontoparietal areas in female subjects (shown in blue in Figure 2 Superior Temporal Gyrus plot), compared to a similar pattern of stronger local connectivity in male subjects (shown in yellow in same plot).

Last, Medial Precentral Gyrus areas (+16,-26,+40) showed relatively higher integration with SN or ventral attention network in female subjects. In contrast, lateral Precentral Gyrus areas (+54,-4,+22) showed higher integration with the same networks in male subjects, while in female subjects this area showed stronger local correlations (shown in blue in Figure 2 Precentral Gyrus plot).

From a validation and generalization's perspective, it is interesting to question whether the same or similar results would have been observed if, instead of using 10 eigenpattern scores, based on a conservative suggestion to maintain approximately a 20:1 ratio between subjects and eigenpatterns, we would have chosen a different number. To that end we repeated the previous group-level analyses evaluating gender differences in connectivity but now using different number of eigenpatterns scores, ranging from 1 to 100, and compared the resulting fc-MVPA statistic parametric maps $F(x)$.

The results (Figure 3 top) show very similar $F(x)$ statistics when varying the number of eigenpatterns around the k=10 value selected for our original analyses. In addition, the distribution of resulting statistics across the entire brain (Figure 3 bottom) shows high sensitivity across the entire range of evaluated k values, consistent with the sensitivity simulations in the sections below, and with average sensitivity peaking at k=50 (approximately a 4:1 ratio in subjects to eigenpatterns) for detecting gender effects in this dataset. While there were several areas like superior Poscentral Gyrus where the statistics peaked at relatively low number of eigenpatterns, suggesting that the effects in these areas may be best represented by the first few fc-MVPA eigenpatterns (i.e. they may be better described in terms of common/large sources of variability across subjects), there was also many areas where the $F(x)$ statistics peaked when using a large number of eigenpatterns (e.g. 50 or above), suggesting that there may still be widespread gender differences in functional connectivity beyond those highlighted in our original analyses and described in Figure 2 that are better expressed in some of the higher-order fc-MVPA eigenpatterns (i.e. representing more subtle or less common sources of inter-subject variability).

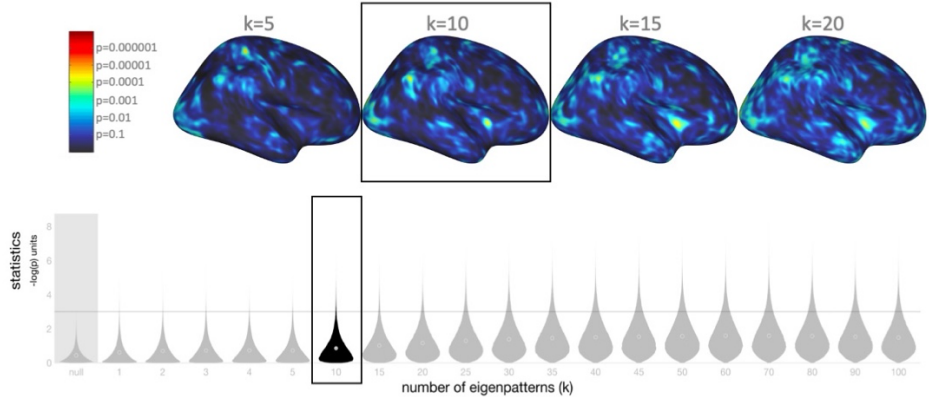

**Figure 3. Selecting different number of fc-MVPA eigenpatterns**. Difference in fc-MVPA statistic parametric maps evaluating gender differences in connectivity, when varying *k*, the number of fc-MVPA eigenpatterns used in the analysis, from k=1 (left) to k=100 (right). For reference, the original results shown in Figure 2 used k=10 (highlighted here inside black box). **Top**: Statistic parametric maps with color coding showing voxel-level -$\log_{10}$(p) values for four different choices of k (from 5 to 20). The results show consistent statistic parametric maps across different *k* values. **Bottom**: Distribution of fc-MVPA statistics across all gray matter voxels with *k* ranging from 1 to 100, compared to null hypothesis distribution (shown in leftmost 'null' histogram). The results indicate high sensitivity across the entire range of evaluated *k* values, with sensitivity peaking at around k=50 (close to a 4:1 ratio in subjects to eigenpatterns) for detecting widespread gender effects in this dataset.

**Fc-MVPA eigenpatterns $P(x)$, interpretation and examples**

In addition to enabling brain-wide connectivity inferences, the fc-MVPA procedure produces a model-free characterization of the observed inter-subject variability in functional connectivity in terms of the resulting eigenpatterns $P(x)$, which can be useful on its own. In this context, the *eigenpatterns*, defined as the columns of the voxel-specific matrix $P(x)$, represent a set of mutually orthogonal spatial patterns, different for each voxel, and best characterizing the diversity across subjects in functional connectivity between this voxel and the rest of the brain. By convention they are sorted in descending order based on the proportion of the total inter-subject covariance explained by each eigenpattern.

In practice, while it is straightforward to compute and store the eigenpattern scores $s_n(x)$, storing the entire set of eigenpatterns $P(x)$ can be particularly demanding as it consists of a set of orthogonal whole-brain maps for each individual voxel. Luckily it is simple to define $P(x)$ analytically at any individual voxel $x$ from its associated eigenpattern scores as:

$$P(x) \propto \sum_n r_n^t(x) \cdot s_n(x) \tag{11}$$

which can be generalized to define characteristic eigenpatterns over small homogeneous areas by integrating the corresponding voxel-specific eigenpatterns:

$$P(\Omega) \propto \sum_n \int_{x \in \Omega} r_n^t(x) \cdot s_n(x) \, dx \qquad (12)$$

Reporting and describing the fc-MVPA eigenpatterns in Equation 12 over a small area, along with the proportion of the total covariance explained by each eigenpattern at this area, allows to gain a better understanding of the main factors affecting inter-subject heterogeneity in functional connectivity between this area and the rest of the brain.

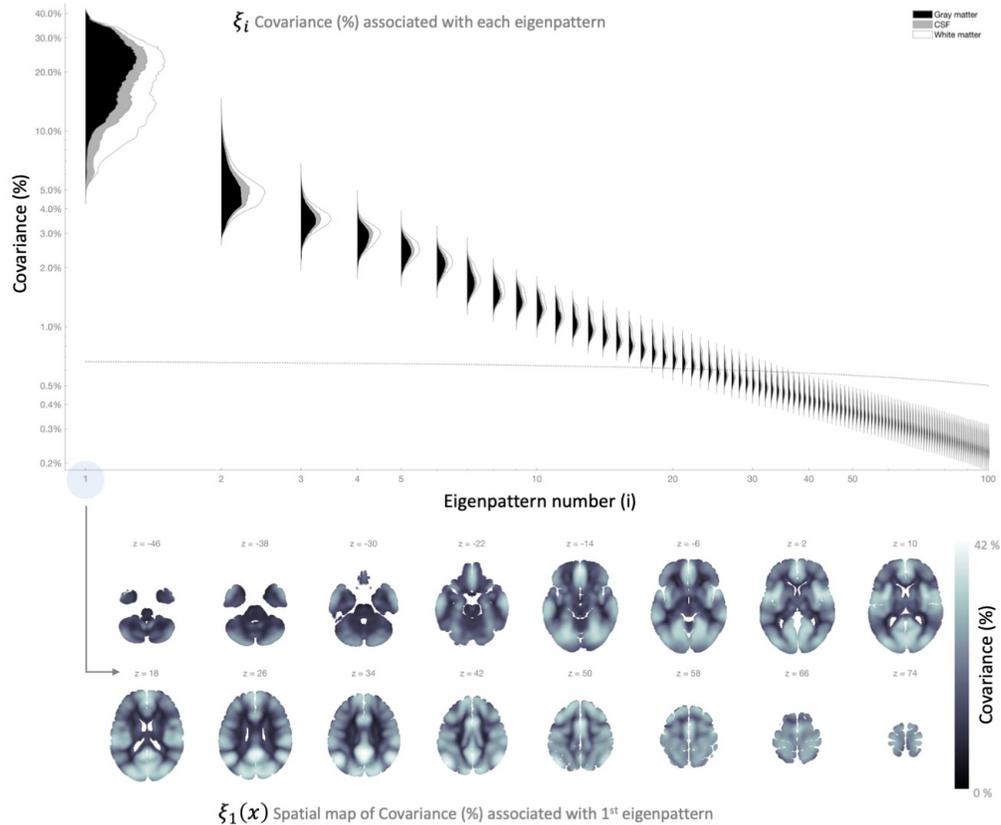

**Figure 4. Percentage of total covariance associated with each fc-MVPA eigenpattern**. **Top**: histogram of $\xi_i(x)|1 \leq i \leq 100$ values, percentage of the total covariance explained by each of the first 100 eigenpatterns. Histrograms are further broken down by the most likely tissue class (gray matter in black, CSF areas in grey, and white matter in white) at each individual voxel as defined by SPM's tissue probability map templates. **Bottom**: spatial map $\xi_1(x)$ showing the proportion of the total inter-subject covariance explained by each voxel's first eigenpattern (a measure of the overall inter-subject homogeneity in functional connectivity patterns at each voxel; see text for details)

For example, from the analysis of the same resting state data of 198 subjects in the Cambridge dataset, the map $\xi_1(x)$ shown in Figure 4 (bottom) describes the proportion of the total inter-subject covariance explained by each voxel's first eigenpattern. In this sample, $\xi_1(x)$ values range between 5% to 42% across different voxels. The values $\xi_1(x)$ associated with the first eigenpattern are particularly interesting because they provide a simple measure of the overall inter-subject homogeneity of the connectivity maps at each voxel (with higher values indicating higher homogeneity), as the first eigenpattern can be often expected to lie in the direction of the average pattern of connectivity with that voxel (as illustrated in the analyses below). Overall, this sample $\xi_1(x)$ map shows high contrast between gray matter areas and other tissue classes, with higher values within gray matter areas, particularly those located along cortical gyra, and relatively lower values for areas located deeper into cortical sulci. Some of the regions that show highest homogeneity include Medial Prefrontal, Posterior Cingulate, or Lateral Parietal areas, part of the Default Mode Network (DMN), as well as anterior Insula and other Salience Network (SN) areas. In contrast, cerebellar, subcortical, and Limbic Network areas are some of the regions that show the most heterogeneous functional connectivity profiles across subjects (lowest $\xi_1(x)$ values).

Looking at the contribution of additional eigenpatterns beyond the first one, Figure 4 (top) shows histograms of the values $\xi_i(x)$, the percentage of the total inter-subject covariance explained by each of the first 100 eigenpatterns. The distribution shows a strongly anisotropic covariance (for reference, the dashed line simulates the expected percent covariance values associated with each eigenpattern if the inter-subject covariance was isotropic, with equal covariance along all subject dimensions and approximate 100-voxel spatial resels). In general, around 20 eigenpatterns are associated with higher-than-average covariance values. The first 5 eigenpatterns combined explain a range between 18% and 51%, 10 eigenpatterns between 28% and 57%, and 20 eigenpatterns between 39% and 64%, of the total inter-subject covariance (from a maximum of 197 eigenpatterns that could be theoretically computed from this sample).

While $\xi_i(x)$ maps allow us to explore how the degree of anisotropy varies across different areas, it is often also of interest to display the actual eigenpatterns $\boldsymbol{P}(x)$ at some particular representative locations, in order to better understand the *shape* of that inter-subject covariance. Figure 5 shows the first 5 eigenpatterns at 14 example locations. These 14 locations were manually chosen to illustrate some of the similarities and differences across different locations in inter-subject variations of functional connectivity patterns. They were selected among the set of all local maxima of the cumulative $\sum_{i=1}^{5}\xi_i(x)$ maps (shown in Figure 5 center image for reference) trying to cover most of the larger clusters observed there. As this figure illustrates, the first eigenpattern across different locations (shown in the leftmost portion of each individual-region display in Figure 5) often reflects a pattern mimicking the average connectivity between each location and the rest of the brain. For example, the first eigenpattern at the Posterior Cingulate gyrus, a region part of the Default Mode Network (DMN), reflects an arrangement very similar to the expected pattern of positive and negative associations with the DMN, and the same arrangement appears in the first eigenpattern at other distant but related locations, such as Frontal Medial Cortex. Similarly, the first eigenpattern at Anterior Insula or Anterior Cingulate also shows similar profiles mimicking Salience or Ventral Attention Network connectivity. In contrast, second- and higher- order eigenpatterns, even from regions that are part of the same network, show noticeable differences in their profiles, possibly indicating non-overlapping sources of inter-subject variability beyond simple within-network connectivity variations. Other regions, in contrast, show eigenpatterns that reflect perhaps competing contributions. For example, in Paracingulate Gyrus the first eigenpattern reflects sources of variability in connectivity with nearby DMN areas while the second eigenpattern possibly reflects variability in connectivity with Anterior Cingulate and Medial Prefrontal regions.

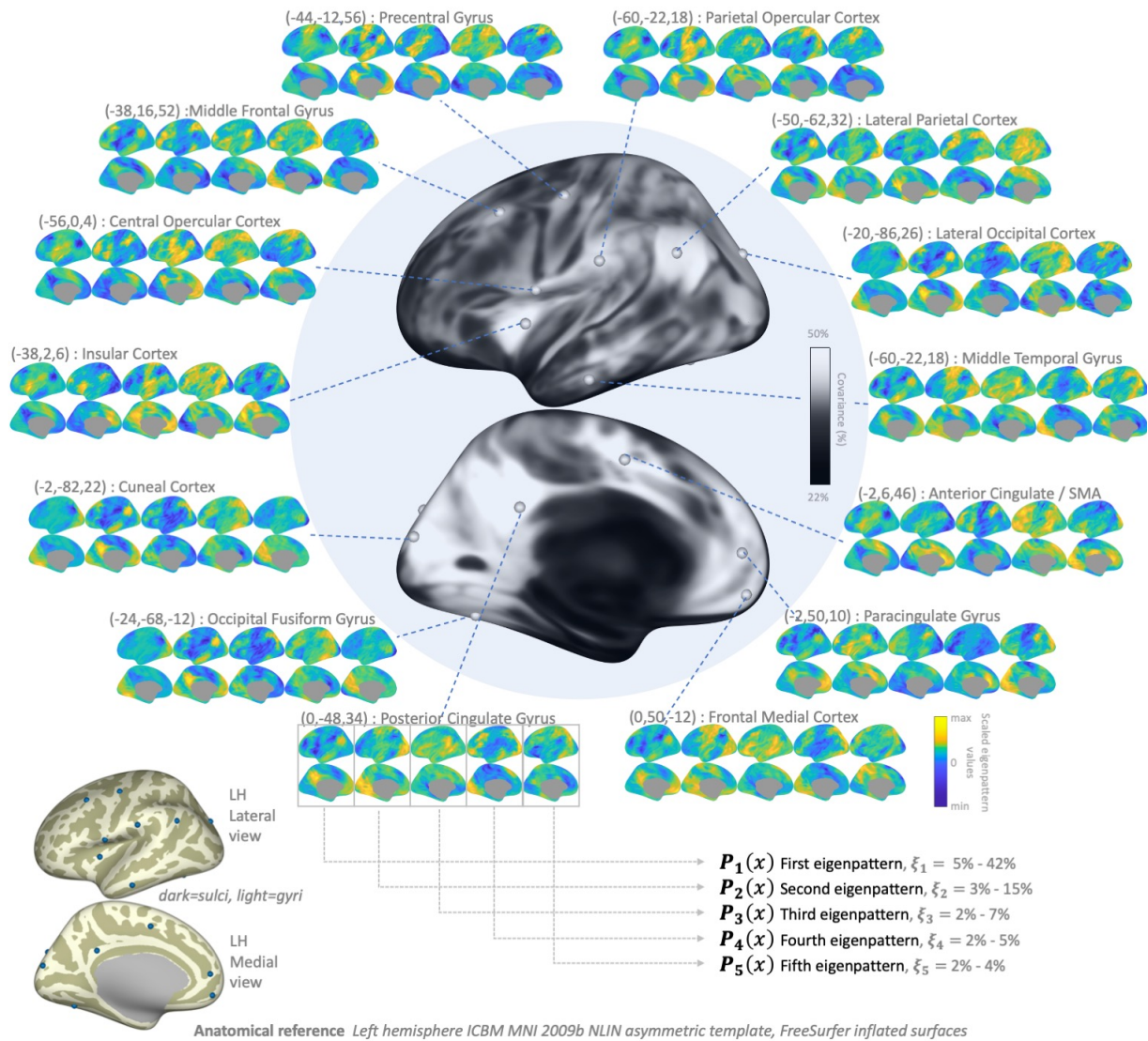

**Figure 5. First 5 fc-MVPA eigenpatterns, characterizing the principal components of the local inter-subject heterogeneity in functional connectivity maps.** The central display shows the cumulative total covariance in functional connectivity patterns explained by the first 5 eigenpatterns at each voxel (colormap values range between 22%/black to 50%/white). The first five eigenpatterns at 14 manually-defined example locations are shown in a circular display. In each of these plots, eigenpatterns range from first/left to fifth/right, and each eigenpattern is shown projected to a left hemisphere lateral (top plot) and medial (bottom plot) views, on a relative color scale ranging from blue (highest negative values for each eigenpattern) to yellow (highest positive values).

From a validation's perspective, it is interesting to note that the covariance explained by the first few fc-MVPA eigenpatterns and represented by the $\xi_i(x)$ values is always, by nature of the fc-MVPA SVD procedure, larger than what could be achieved by any other alternatively-defined eigenpatterns. In particular, it is larger than the spatial patterns that would result from a standard ICA or PCA decomposition of the same functional data. In order to highlight this, we computed on this dataset a Principal Component Analysis in CONN using the same concatenative approach and dimensionality reduction steps as in Calhoun et al. 2001 to produce a set of representative components sorted by decreasing explanatory power (shown in Figure 6 bottom). We then computed, for each of these components, the covariance in functional connectivity with each individual voxel along those dimensions, and plotted histograms of the resulting cumulative variance as a function of the number of components retained (Figure 6 top). As expected, the covariance explained cumulatively by the first *k* fc-MVPA eigenpatterns at each individual voxel (shown in gray in Figure 6 top) is always equal or larger than the covariance explained cumulatively by the same number of PCA components (shown in black in the same plots). While this is a necessary consequence of the SVD properties as used in the context of fc-MVPA, it is important to note that in particular this implies that, if we would like to characterize the functional connectivity pattern at each voxel using a reduced fixed number of scores, then the representation produced by the fc-MVPA eigenpattern scores would always be more efficient (it would better approximate the functional connectivity data) than a equally-sized multivariate representation produced by characterizing each voxel connectivity in terms of network-level properties (at least for the general class of linear transformations projecting each connectivity pattern onto multiple networks, including those resulting from Principal Component or Independent Component Analyses of the same data). This, naturally, also supports the use of fc-MVPA eigenpattern scores in the context of brain-wide connectome inferences as a rich low-dimensional representation of the functional connectivity patterns for each subject.

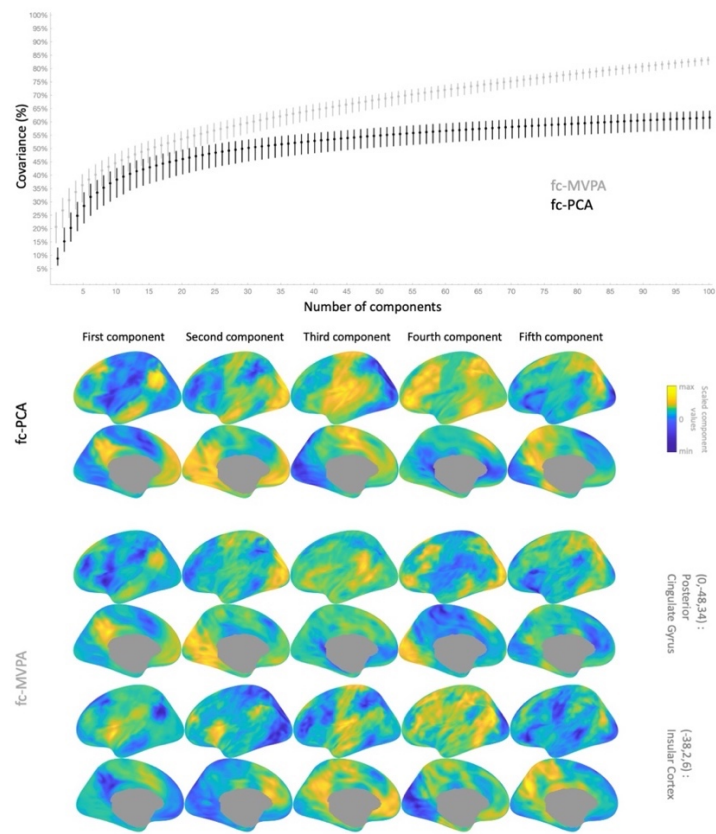

**Figure 6. Comparison between PCA and MVPA components**. **Top**: Median (dots) and 25%-75% percentile range (vertical lines) of the total covariance in functional connectivity patterns at each voxel explained cumulatively by the first *k* components from a functional connectivity Principal Component Analysis (black dots and lines), and by the first *k* fc-MVPA eigenpatterns (light gray dots and lines), from the analysis of the same sample dataset (Cambridge, n=198 dataset). **Bottom**: First five principal components from PCA (first row) and from fc-MVPA (second and third row, first five eigenvariates shown only at two sample locations: posterior cingulate and anterior insula). Each row shows individual components sorted from first/left to fifth/right, projected to a left hemisphere lateral view (top image) and medial view (bottom image), on a relative color scale ranging from blue (highest negative values for each component) to yellow (highest positive values). Larger explanatory power of fc-MVPA components compared to PCA (shown on top figure) stems largely from the ability of fc-MVPA components to adapt to the specificity of the functional connectivity patterns at each individual location (as exemplified in the bottom figures by the differences and commonalities between the components describing posterior cingulate vs. anterior insula connectivity patterns)

**Simulations: validity and sensitivity of fc-MVPA statistics**

In order to evaluate the validity and sensitivity of the general fc-MVPA inferential approach we constructed a set of simplified simulations. All of the simulations consider a dataset with 50 subjects. Each subject's BOLD data encompasses 50 timepoints and 1,000 voxels. For each voxel, the simulated BOLD timeseries contained a mixture of noise (independent samples from a Gaussian distribution for each timepoint, computed separately for each voxel and subject, and spatially convolved with a Gaussian filter with FWHM 10 voxels) and signal (independent samples from a Gaussian distribution for each timepoint, computed separately for each subject and shared across all voxels where the signal was present). The signal was only present in one half of the subjects, and, among those subjects, only within 10% of all contiguous voxels. For each individual simulation Equation 5 was used to estimate the projection matrix P(x) at each individual voxel, and Equations 7 & 8 were used to evaluate between-group differences in the patterns of connectivity between this voxel and all other voxels, using a 50-by-2 design matrix **G** characterizing the two groups of subjects and a between-subjects contrast vector **C** = [-1 1] evaluating the difference in functional connectivity between the two groups.

For each individual simulation we computed the statistical parameter map *F(x)* and the associated map of raw/uncorrected voxel-level p-values evaluating the null hypothesis separately for two different voxels: one where the signal was present (so the connectivity between that voxel and all of the other voxels is expected to differ between the two subject groups), and one where the signal was not present (so the connectivity between that voxel and all other voxels is not expected to differ between the two subject groups). We run 10,000 simulations, and from their results computed summary Receiver Operating Characteristic (ROC) curves describing the true positive rate (probability of a voxel showing a significant between-group difference in connectivity) as a function of different prescribed false positive rates (p-value threshold used to determine significance), for each of these two representative voxels. The results from the first voxel, where the signal was present, were used to obtain estimates of the sensitivity of voxel-level fc-MVPA connectome inferences (*sensitivity* analyses), and the results from the second voxel, where the signal was not present, were used to obtain estimates of the validity of this inferential procedure (*validity* analyses).

Each of the above sets of 10,000 simulations were repeated 50 times, each time using a different number of eigenpatterns retained (ranging between 1 and 50) in Equation 5. In addition, all of the above simulations were repeated under six different scenarios in order to further evaluate the robustness of the obtained sensitivity and validity estimates in the presence of: a) varying levels of spatial cross-correlation of BOLD noise (FWHM set to 10, and 100 voxels); b) varying number of timepoints in BOLD scanning sessions (10, and 100 samples); and c) varying number of subjects in the study (10, and 100 subjects).

The results of the *validity* analyses are shown in Figure 7. The reported voxel-level p-values (shown in the x-axis labeled as *false positive rate*) from fc-MVPA inferences matched very precisely the empirically observed false positive rates (shown in z-axis labeled as *positive rate*), with all tested conditions showing accurate diagonal ROC curves. Differences between reported voxel-level p- values and observed false positive rates were below ± 0.22% in 50% of all simulations, and below ± 0.98% in 99% of all simulations. When controlling voxel-level false positives at a 5% level, and across a total of 386 sets of different conditions evaluated, the empirical false positive rate observed across the 10,000 fc-MVPA analyses within each set ranged between 4.5% and 5.4% (Figure 7 top right). Statistics remained valid across the entire range of eigenpatterns tested up to the point where the number of eigenpatterns (*a* in Equation 8) equals the error degrees of freedom (*b* in Equation 8, equal to the number of subjects minus the number of model regressors, or 48 in our simulations) where the data covariance becomes rank deficient and the likelihood ratio test assumptions no longer hold.

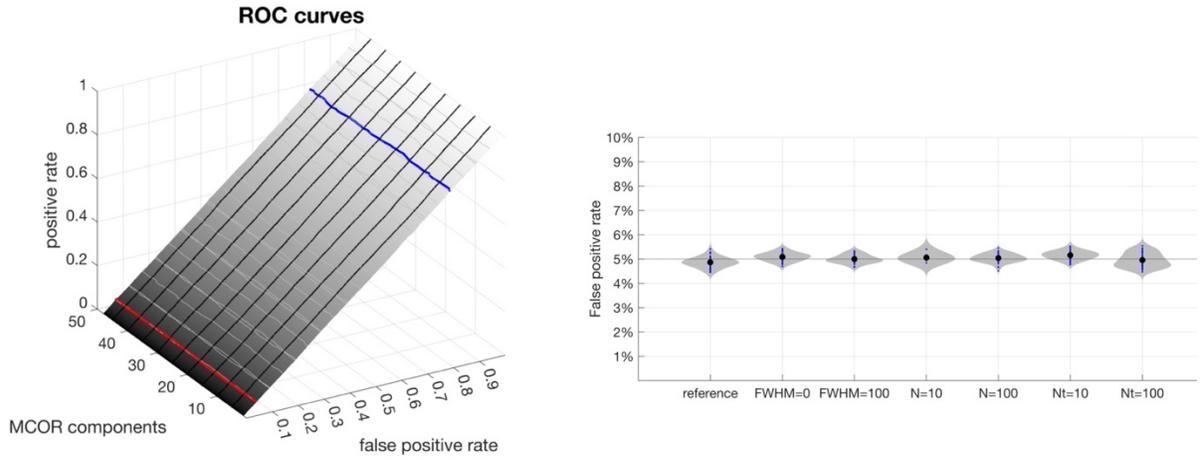
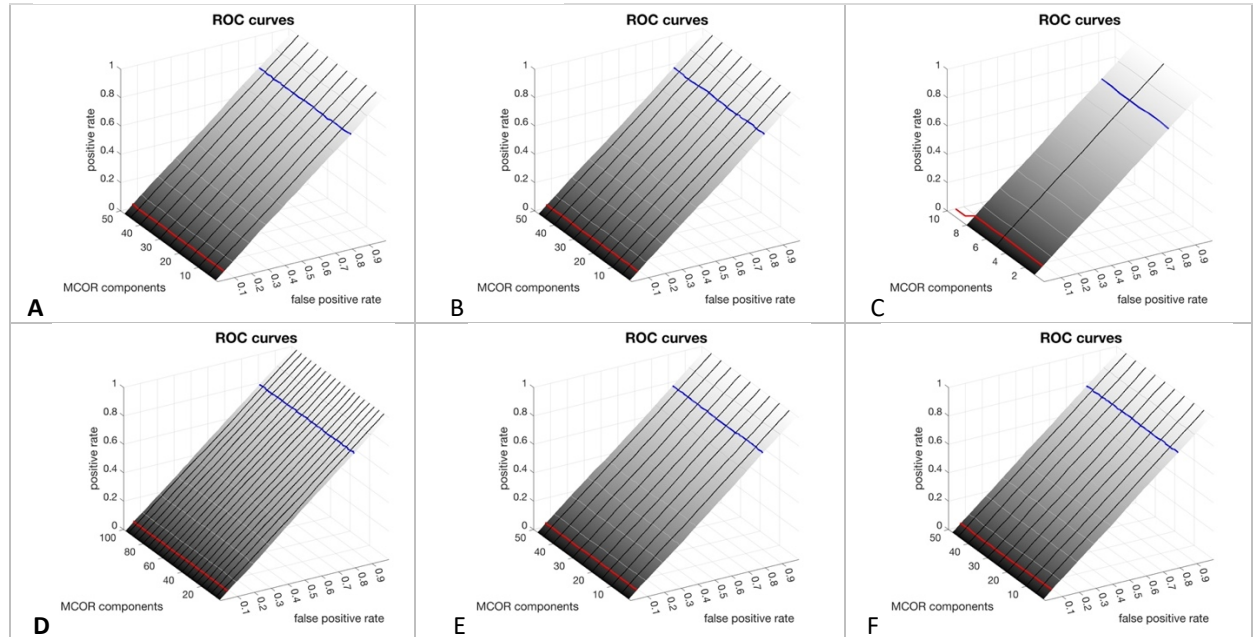

**Figure 7. Validation of fc-MVPA voxel-level inferences**. Analysis of Receiver Operating Characteristic curves evaluating between-group differences in functional connectivity under the null hypothesis (when there are no true differences in the population). **Top left**: surfaces, and highlighted thick black lines, show, for a chosen combination of false positive threshold (*false positive rate* x-axis) and number of eigenpatterns (*MCOR components* y-axis), the resulting proportion of false positive results (*positive rate* z-axis), where the fc-MVPA procedure would falsely conclude there is a significant difference in connectivity between the groups. The red line marks the observed rate of false positives when fixing the prescribed false positive rate threshold at a fixed 5% level (graphically, the intersection of each ROC surface and a vertical plane with constant false positive rate = 0.05), matching the expected 5% level, while the blue line marks the minimum false positive threshold needed to reach a sensitivity of 80% or higher (graphically, the intersection of ROC surfaces and a horizontal plane with constant positive rate = 0.80). **Top Right**: Observed false positive rates (y-axis) when using fc-MVPA statistical analyses controlled at a $p<.05$ level across the reference simulations ('reference') and simulations evaluating different conditions (FWHM=0, FHWM-100, N=10, N=100, Nt=10, Nt=100). The average (black dots) and histogram (gray surfaces) of the observed false positive rates across these simulations all indicate an appropriate match to the expected/prescribed false positive level (5%). **Bottom**: evaluating validity under different conditions: (**A**) low spatial autocorrelation (FWHM=0); (**B**) large spatial autocorrelation (FWHM=100 voxels); (**C**) low number of subjects (N=10); (**D**) high number of subjects (N=100); (**E**) short scanning session (Nt=10); (**F**) long scanning session (Nt=100)

The results of the *sensitivity* analyses are shown in Figures 8. Generally, sensitivity was large across the entire range of eigenpatterns tested, only decreasing markedly as the number of eigenpatterns approached their maximum possible value. For example, sensitivity at a $p<.05$ level was above 80% in the main simulations (with 50 subjects) as long as the number of eigenpatterns was kept below 42, below 92 in the simulations with 100 subjects, and below

4 in the simulations with 10 subjects. While optimal sensitivity will naturally vary on multiple factors, including the size and nature of the effects that we are trying to evaluate, several trends in sensitivity were apparent from different scenarios evaluated. In particular, the degree of spatial autocorrelation in the functional data (simulations A-B in Figure 8) appeared to have a relatively small effect on sensitivity, while the number of subjects in the study (simulations C-D) and the duration of the scanning session (simulations E-F) both had a larger impact. For example, when fixing the number of eigenvariates to 5, sensitivity to detect a group effect was above 99% at a p<.05 level in the simulations with 50 or 100 subjects, but sensitivity dropped to 56% when the number of subjects was only 10. Similarly, sensitivity at a p<.05 level was above 99% when the number of simulated timepoints (or equivalently, the number of effective degrees of freedom of the BOLD timeseries in a study after the denoising and band-pass filtering procedure) was above 50, but it dropped to 92% when the number of simulated timepoints was only 10.

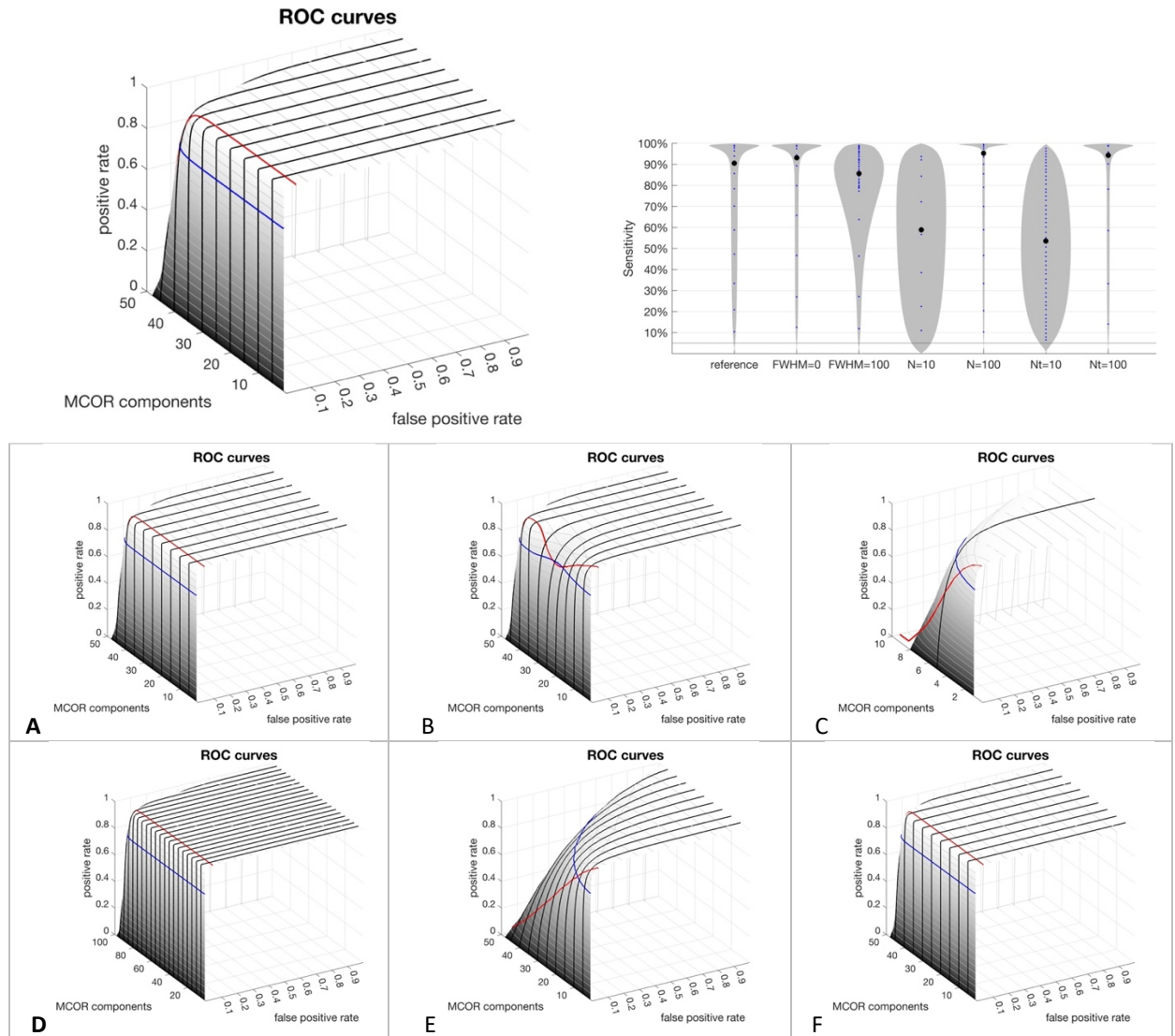

**Figure 8. Sensitivity of fc-MVPA voxel-level inferences**. Analysis of Receiver Operating Characteristic curves evaluating between-group differences in functional connectivity. **Top Left**: surfaces, and highlighted thick black lines, show, for a chosen combination of false positive threshold (*false positive rate* x-axis) and number of eigenpatterns (*MCOR components* y-axis), the resulting proportion of true positive results (*positive rate* z-axis), where the fc-MVPA procedure would correctly conclude there is a significant difference in connectivity between the groups in our reference simulations. **Top Righ**t: Observed true positive rates (y-axis) when using fc-MVPA statistical analyses controlled at a p<.05 level across the reference simulations ('reference') and simulations evaluating different conditions (FWHM=0, FHWM-100, N=10, N=100, Nt=10, Nt=100). The average (black dots) and histogram (gray surfaces) of the observed true positive rates, or proportion of significant results, across

these simulations indicate that sensitivity is typically higher when using low or intermediate numbers of eigenpatterns, with poorer sensitivity when the number of timepoints for functional connectivity estimation is low (Nt=10), or when the number of subjects included in the analysis is low (N=10). **Bottom**: evaluating sensitivity under different conditions: (**A**) no spatial autocorrelation (FWHM=0); (**B**) large spatial autocorrelation (FWHM=100 voxels); (**C**) low number of subjects (N=10); (**D**) high number of subjects (N=100); (**E**) short scanning session (Nt=10); (**F**) long scanning session (Nt=100)

**Conclusions**

This manuscript presented the theory and motivation behind functional connectivity Multivariate Pattern Analyses (fc-MVPA), both in the context of brain-wide connectome inferences, as well as a model-free characterization of the heterogeneity in functional connectivity across subjects.

Monte Carlo simulations showed that fc-MVPA inferences remain valid for the entire range of evaluated scenarios, including using any arbitrary number of eigenpattern scores, different sample sizes, and scanning session durations. Simulations and example analyses of gender-related differences in functional connectivity illustrated high sensitivity of fc-MVPA inferential statistics to detect meaningful effects across the entire human connectome. In addition, an example analysis of fc-MVPA eigenpatterns in functional connectivity during resting state showed rich and varied sources of inter-subject heterogeneity in functional connectivity.

One of the main practical advantages of fc-MVPA procedure in the context of brain-wide connectome inferences, is that it combines the benefits of pattern analysis techniques, such as the increased interpretability and reduced noise of lower-dimensional projections, with the benefits of a classical statistical framework, such as the ability to use popular techniques for multiple comparison control (e.g. TFCE), or well understood statistical control procedures when dealing with potential confounders (e.g. ANCOVA in this manuscript example analyses). Similarly, fc-MVPA eigenpattern representations offer a natural way to extend common multidimensional reduction approaches in neuroimaging, such as ICA or PCA, to begin considering the specificity of these lower-dimensional representations across different brain areas. From its theoretical and practical advantages, we believe that fc-MVPA can be a powerful and hopefully useful tool for researchers to further explore the complexities of the human connectome.

**Conflict of Interest**

The authors have no real or potential conflicts of interest to declare that are relevant to the content of this article.

**Acknowledgements**

This research was supported by the National Institute on Deafness and Other Communication Disorders (R01 DC007683, R01 DC002852, R01 DC016270), National Institute of Neurological Disorders and Stroke (U01 NS117836) and National Institute of Mental Health (U01 MH108168)

**Appendix I. Efficient computation of multivariate patterns**

A diagram of the computational procedure followed to compute eigenpattern scores is shown in Figure A.I.1.

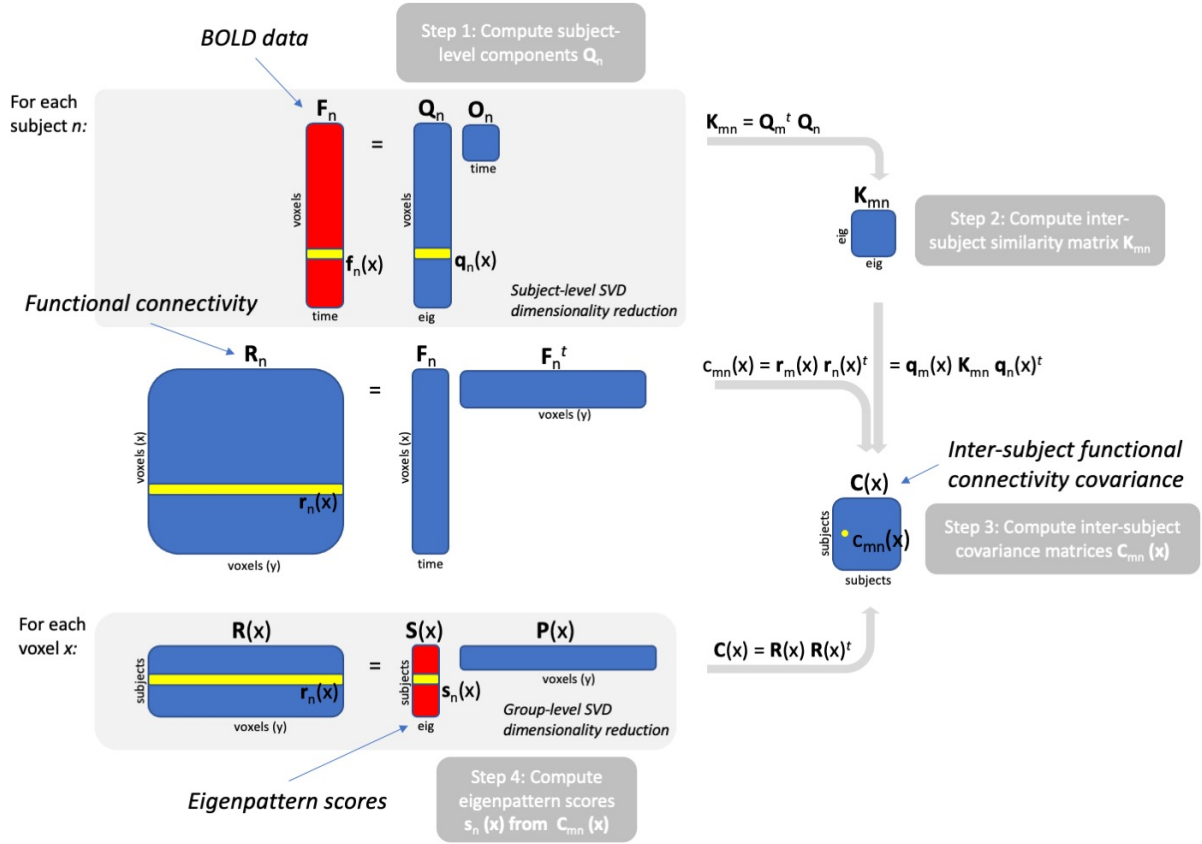

**Figure A.I.1. Efficient computation of eigenpattern scores.** Schematic summarizing the steps involved in computing eigenpattern scores for each subject. First, the procedure loops over each subject to decompose the voxels-by-time BOLD data matrix using Singular Value Decomposition (SVD) into orthogonal factors $Q_n$. Them the inter-subject similarity between these orthogonal factors is computed. Last, the procedure loops over each voxel to compute: a) the subject-by-subject functional connectivity covariance matrix C(x); and b) the associated eigenpattern scores by decomposing this covariance matrix using SVD.

Following the general voxel-to-voxel analysis procedure described in Whitfield-Gabrieli and Nieto-Castanon 2012, we first characterize separately for each subject the voxel-to-voxel correlations $r_n(x, y)$ (one voxels-by-voxels matrix for each subject) without loss of generality or precision using a reduced set of orthogonal (not orthonormal) components $q_{n,i}(x)$, with one volume per subject and component, defined from a Singular Value Decomposition (SVD) of the preprocessed and normalized BOLD timeseries $\tilde{f}(x, t)$:

$$q_{n,m}(x) \mid \quad r_n(x, y) \equiv \sum_{t=1}^{T} \tilde{f}_n(x, t) \cdot \tilde{f}_n(y, t) = \sum_{m=1}^{M} q_{n,m}(x) \cdot q_{n,m}(y)$$

(note: $x$ and $y$ index over voxels, while $t$, $i$, and $n$ index over scans, components, and subjects, respectively). The number of components $M$ necessary to fully characterize (without any dimensionality reduction) the voxel-to-voxel correlation matrix is determined by the effective degrees of freedom of the original timeseries (approximately equal to the number of scans minus the number of regressed components multiplied by the proportion of the frequency spectrum covered by the band-pass filter preprocessing step), so the resulting components of the voxel-to-voxel covariance can always be stored using only a portion of the storage necessary to store the original BOLD timeseries.

We also compute from these components and store the cross-covariance matrix **K**, characterizing the spatial similarity between the $m_1$-th SVD component from subject $n_1$ and the $m_2$-th component from subject $n_2$:

$$K_{n_1,n_2,m_1,m_2} \equiv \int q_{n_1,m_1}(x) \cdot q_{n_2,m_2}(x) \, dx \qquad (A1.1)$$

as well as the spatial-covariance matrix **L**, characterizing the degree of smoothness of these components:

$$L_{i,j,n_1,n_2} = \sum_{m_1,m_2=1}^{M} K_{n_1,n_2,m_1,m_2} \cdot \int \frac{\partial}{\partial x_i} q_{n_1,m_1}(x) \cdot \frac{\partial}{\partial x_j} q_{n_2,m_2}(x) \, dx \qquad (A1.2)$$

After computing $K_{n_1,n_2}$ and $q_{n,m}(x)$ for all subjects, we can then easily compute, from these measures alone and separately for each voxel, the voxel-specific subjects-by-subjects cross-covariance $C_{n_1,n_2}(x) \equiv \int r_{n_1}(x,y) \cdot r_{n_2}(x,y) \, dy$, characterizing the between-subjects variability in functional connectivity maps between this voxel and the rest of the brain, as:

$$C_{n_1,n_2}(x) \equiv r_{n_1}(x) \cdot r_{n_2}^t(x) = \sum_{m_1,m_2=1}^{M} q_{n_1,m_1}(x) \cdot q_{n_2,m_2}(x) \cdot K_{n_1,n_2,m_1,m_2} \qquad (A1.3)$$

From the definition of **C(x)** above and Equation 5, it follows that:

$$C(x) = R(x) \cdot R(x)^t = S(x) \cdot D^2(x) \cdot S^t(x) \qquad (A1.4)$$

This means that we can directly compute **S** and **D** as the eigenvectors and squared root of eigenvalues, respectively, of the semidefinite positive matrix **C(x)**, and from this the eigenpattern scores:

$$s_n(x) = e_n \cdot R(x) \cdot P(x) \cdot D^{-1}(x) = e_n \cdot S(x) \qquad (A1.5)$$

where $e_n$ is a unit N-element vector with 1 in the n-th position and 0's otherwise. Overall, this procedure means that we can compute the desired eigenpattern scores $s_n(x)$ at a particular location x directly from an eigenvector decomposition of **C(x)** (Equation A1.4), which, in turn, can be computed only from $K_{n_1,n_2}$ and $q_{n,i}(x)$ (Equation A1.3), without ever having to undertake the considerably more time-consuming operations involved in explicitly computing or storing $R(x)$ or $P(x)$.

The computation of $C(x)$ and $s_n(x)$ is then iterated for each voxel x to create the full set of component volumes (one file per subject and component) storing the eigenpattern scores $s_n(x)$. Note that the number of operations involved in these computations for any given voxel is independent of the number of voxels, and hence the total number of operations will scale only linearly with the total number of voxels in the analysis, as opposed to quadratically as in the original formulation, making the computation of the MVPA eigenpattern scores a considerably faster procedure.

**Appendix II. Preprocessing and analysis of resting state data**

We performed fc-MVPA analyses to identify gender-related differences in functional connectivity using a publicly available dataset (Cambridge 1000-connectomes dataset, n=198; available at NITRC *fcon_1000*). All analyses are performed using CONN21 and SPM12. The functional data was realigned using SPM12 realign & unwarp procedure (Anderson et al. 2001), where all scans are coregistered and resampled to a reference image (first scan of the first session) using b-spline interpolation. Temporal misalignment between different slices of the functional data was corrected using SPM12 slice-timing correction (STC) procedure (Henson et al. 1999). Potential outlier scans were identified using ART (Whitfield-Gabrieli et al. 2011) as acquisitions with framewise displacement above 0.5mm or global BOLD signal changes above 3 standard deviations (Power et al. 2014). Functional and anatomical data were normalized into standard MNI space and segmented into grey matter, white matter, and CSF tissue classes using SPM12 unified segmentation and normalization procedure (Ashburner and Friston, 2005). Last, functional data was smoothed using spatial convolution with a Gaussian kernel of 8mm full width half maximum (FWHM). In addition, default CONN denoising steps were performed including the regression of session effects and their first order derivatives (2 factors), motion parameters and their first order derivatives (12 factors), outlier scans (below 30 components), white matter (5 components), and CSF timeseries (5 components) as potential confounding effects, detrending, and band-pass filtering between 0.008Hz and 0.09Hz (Friston et al. 1996, Behzadi et al. 2007, Chai et al. 2012, Nieto-Castanon, 2020). After outlier removal there were an average of 114.6 functional scans/acquisitions per subject (ranging between 89 to 119), and after denoising and band-pass filtering the residual effective degrees of freedom averaged 43.6 (ranging between 31.0 to 45.8). Residual inter-scan motion (also computed after outlier removal) was 0.087mm on average, with 0.012 mm higher motion (T(196) = 2.75, p = 0.006565) in male subjects (N=75) compared to female subjects (N=123).

Quality control plots were created using CONN based on the analysis of functional connectivity values (Pearson correlation coefficients between the BOLD before and after denoising at each pair of nodes) in a graph formed by 1,000 nodes randomly selected within gray matter voxels. Functional connectivity (FC) distributions indicated reasonably centered distribution of voxel-to-voxel correlation values after denoising (see Figure A.II.1 bottom plot), with global correlation, the average of all correlations between pairs of voxels separately for each subject, equal to GCOR = 0.029 ±0.012 (average ± standard deviation across subjects), which were as expected considerably lower and more stable across subjects compared to global correlations before denoising GCOR = 0.467 ±0.099. Additional QC-FC quality control analyses (shown in Figure A.II.2) computed the estimated strength of residual inter-subject correlations between measures of subject motion (average framewise displacement) and functional connectivity strength at each edge within the same random graphs (Ciric et al. 2017). These plots also indicate appropriate levels of denoising, with distributions of motion-connectivity correlations after denoising (gray area in Figure A.II.2 bottom plot) similar -95.6% match- to those expected by chance (red dashed lines). Similarly, QC-FC associations between functional connectivity strength and the number of valid scans remaining for each subject after outlier detection showed negligible effects after denoising, with 97.2% match with the null hypothesis distribution.

First-level analyses included fc-MVPA, with subject-level dimensionality reduction set to 64 dimensions, and the number of eigenpatterns estimated at each voxel set to 100. Second-level analyses then used the first 10 components among the 100 estimated eigenpatterns in order to maintain a conservative 20:1 ratio of subjects-to-components. The resulting eigenpattern score volumes were entered into a multivariate second-level General Linear Model analysis with 198 samples (subjects) and 10 observations (eigenpattern scores). The model included two factors: gender (categorical factor with two levels) and subject motion (a continuous factor, encoding residual average framewise displacement after removal of outlier scans for each subject). The contrast [-1 1 0] testing gender-related differences in connectivity controlling for subject motion was evaluated using F(10,186) statistics. Threshold Free Cluster Enhancement scores (Smith and Nichols, 2009) were estimated using CONN's default values for F-statistics H=1, E=0.5, and H$_{min}$=1. Non-parametric randomization test (Bullmore et al. 1999) with 1,000 simulations were used to determine TFCE score statistics under the null hypotheses. Analysis results were thresholded at a family-wise error corrected p-FWE < 0.05 level. Last, for effect-size estimation we computed $h_{\text{map}}$ using Equation 9 for each significant cluster, characterizing the difference in connectivity between male and female subjects when compared at the same level of the control covariate (subject motion).

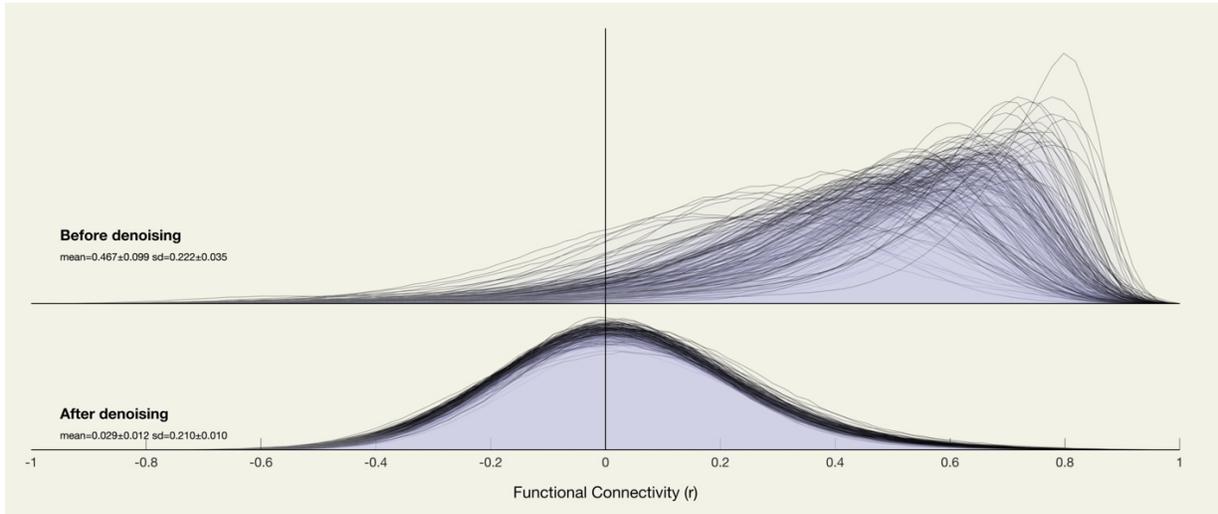

**Figure A.II.1. Quality control plots of functional connectivity (FC) histograms**. Distributions of Pearson correlation values between the BOLD signal at 1,000 random nodes within gray matter voxels, computed separately for each subject before denoising (top plots) and after denoising (bottom plots). Distributions after denoising appeared well centered (distributions mode close to r=0 values) and homogeneous (distributions shape similar across subjects), both markers indicative of appropriate levels of denoising of the BOLD signal.

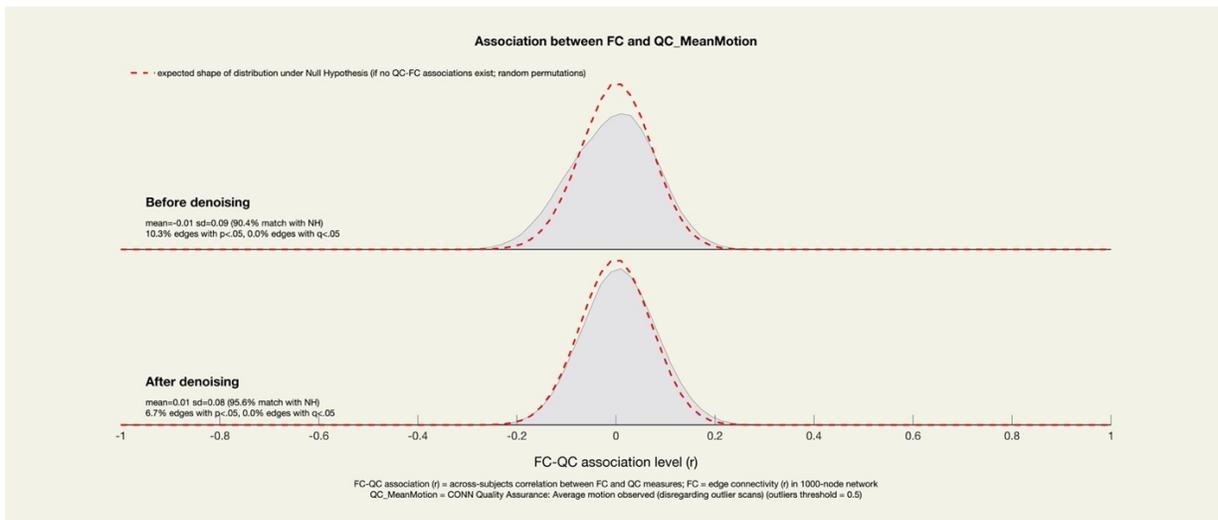

**Figure A.II.2. Quality control plots of QC-FC correlations**. Distributions of Pearson inter-subject correlation values between the functional connectivity strength at all edges in a 1000-node graph and a measure of subject motion (average framewise displacement), computed separately before denoising (top plots) and after denoising (bottom plots). Red dashed lines show the expected distribution of QC-FC correlations in the absence of meaningful associations between connectivity strength and subject motion (obtained using permutation analyses of the same data). QC-FC distributions show a good match with the expected distribution under the null hypothesis (95.6% overlap between the two distributions) after denoising, with percent match above 95% indicative of appropriate levels of denoising of the BOLD signal (Nieto-Castanon 2020).